# Imaging an aligned polyatomic molecule with laser-induced electron diffraction

Michael Pullen[1,*], Benjamin Wolter[1,*], Anh-Thu Le[2], Matthias Baudisch[1], Michaël Hemmer[1], Arne Senftleben[3], Claus Dieter Schröter[4], Joachim Ullrich[4,5], Robert Moshammer[4], Chii-Dong Lin[2], Jens Biegert[1,6]

[1]ICFO-Institut de Ciencies Fotoniques, Mediterranean Technology Park, 08860 Castelldefels (Barcelona), Spain

[2]J. R. Macdonald Laboratory, Physics Department, Kansas State University, Manhattan, Kansas 66506-2604, USA

[3]Universität Kassel, Institut für Physik und CINSaT, Heinrich-Plett-Str. 40, 34132 Kassel, Germany

[4]Max-Planck-Institut für Kernphysik, Saupfercheckweg 1, 69117 Heidelberg, Germany

[5]Physikalisch-Technische Bundesanstalt (PTB), Bundesallee 100, 38116 Braunschweig, Germany

[6]ICREA-Institució Catalana de Recerca i Estudis Avançats, 08010 Barcelona, Spain

*authors contributed equally to this work

**Laser-induced electron diffraction is an evolving tabletop method, which aims to image ultrafast structural changes in gas-phase polyatomic molecules with sub-Ångström spatial and femtosecond temporal resolution. Here, we provide the general foundation for the retrieval of multiple bond lengths from a polyatomic molecule by simultaneously measuring the C-C and C-H bond lengths in aligned acetylene. Our approach takes the method beyond the hitherto achieved imaging of simple diatomic molecules and is based upon the combination of a 160 kHz mid-IR few-cycle laser source with full three-dimensional electron-ion coincidence detection. Our technique provides an accessible and robust route towards imaging ultrafast processes in complex gas phase molecules with atto- to femto-second temporal resolution.**

Dynamic imaging of chemical reactions or biological functions is one of the grand challenges of science[1,2]. These processes are typically triggered by sub-Ångström scale events that are initiated on the few femtosecond timescale. An imaging method that could achieve the required spatio-temporal resolutions would provide revolutionary insight into the connection between molecular structure at critical transition points and barrier heights; hallmark examples are transition states[3], rapid dynamics in the vicinity of conical intersections[4], or proton migration and isomerisation[5]. The capability of imaging the motions of the atomic constituents during these processes with few-femtosecond temporal and sub-Ångström spatial resolutions therefore represents a paradigm shift in modern physics and chemistry.

Ultrafast electron diffraction (UED) is capable of resolving atomic positions with sub-Ångström resolution[6], however, the achievable temporal resolution is currently limited to hundreds of femtoseconds mainly due to Coulomb repulsion in the electron bunch. Such temporal resolution is not sufficient to resolve the initiation reactions and ultrafast changes of the prototypical processes mentioned above. Current developments therefore aim at reducing space charge[7] or using relativistic electron bunches[8]. X-ray diffraction methods[9] currently suffer from spectro-temporal jitter and are only available at large-scale facilities. These restraints have motivated the development of new dynamical imaging techniques, largely for the gas phase, such as chirped encoded recollisions[10], photoelectron holography[11], femtosecond photoelectron diffraction[12], Coulomb explosion imaging[13] and laser assisted electron diffraction[14].

Clearly, versatile laboratory scale table-top methods that provide the combined spatial and temporal resolutions would signify a breakthrough, especially for the imaging of gas-phase molecular dynamics. Laser-induced electron diffraction (LIED) is such a method and is based on probing an objects' structure using its own electrons that are rescattered during strong-field induced recollisions[15-17]; this process is depicted in Fig. 1 where the longitudinal and transverse momenta are defined as $k_{\parallel} = k_y$ and $k_{\perp} = \sqrt{k_x^2 + k_z^2}$, respectively. Coherent sub-cycle elastic scattering of the electron wavepacket with attosecond (single pulse) to femtosecond (pulse train) resolution retains structural information of the ionic species in the resultant diffraction pattern[18-21]. The challenge lies in *simultaneously* fulfilling the extremely stringent conditions for LIED in order to extract structural information; these are: (i) achieving high recollision energies despite a small fraction of target ionisation, (ii) achieving core penetrating collisions and sufficient momentum transfer with the scattered electron, (iii) driving recollision in the quasi-static regime to enable extraction of field-free diffraction images from the photoelectron momentum spectra. When these conditions are met, the method of molecular structure retrieval is similar to conventional electron or X-ray diffraction, with the added benefit of femtosecond temporal resolution of the driving laser. In general, state of the art near-infrared lasers cannot fulfill these combined conditions, except in the specific case of homonuclear diatomic molecules such as $O_2$[19]. Recently, these conditions were satisfied with ~2 μm lasers[22,23] and structural retrieval of $N_2$ and $O_2$ molecules was demonstrated. Spatial resolutions of 0.05 Å were reported, which were sufficient to image a 0.1 Å contraction of the simple $O_2$ molecule during the ~5 fs it takes an electron to rescatter. This result established the potential of LIED as a dynamical imaging technique with sub-Ångström spatial and few-femtosecond temporal resolutions.

To harness the combined temporal and spatial resolutions of LIED and applying it to polyatomic molecules (i.e. systems with three or more atoms that exhibit full prototypical molecular dynamics) requires addressing a decisive and unresolved issue, namely the fact that launching the recollision

(imaging) electron initiates molecular distortion and eventually fragmentation. Therefore, a certain portion of the detected electrons present an unwanted background that can make imaging difficult or even impossible. We resolve this problem through ion-electron coincidence detection and the retrieval of the doubly differential cross section. This ensures unambiguous imaging of the molecular structure, or fragments, of interest. In addition to this major concern, there are other experimental obstacles that must be overcome: Firstly, complex molecules commonly have ionisation energies around and below 10 eV, which necessitates the use of mid-IR driving lasers in order to avoid ionisation saturation. Mid-IR sources also have the added benefit that electrons with the required energies are liberated at lower intensities, which results in less distortion of the molecule. Secondly, because each constituent atom has a unique scattering cross section, a careful selection of the electron scattering parameters ensures that they all contribute significantly to the scattering and hence facilitates the simultaneous determination of multiple bond lengths. Thirdly, to resolve the increased structural complexity, it is highly beneficial for the gas ensemble to be anisotropically distributed with respect to the molecular axis in order to remove averaging effects[20].

Here, we meet all of these challenges through a combination of experimental methodologies: A unique home-built optical parametric chirped pulse amplification (OPCPA) source provides 1.7 μm and 3.1 μm pulses at a repetition rate of 160 kHz[24] and with excellent long-term stability. The 1.7 μm light is used to impulsively align the target molecule while the 3.1 μm light induces electron rescattering. The lower efficiency of the rescattering process[25,26] at longer wavelengths is more than compensated for by the two orders of magnitude higher repetition rate of our source compared to typical 1 kHz systems. Equally as important is the reaction microscope (ReMi) detection system that allows a careful selection of the relevant channels (over both electron energies and scattering angles) from the doubly differential cross-section in coincidence[27].

We validate our unique experimental approach by imaging the aligned polyatomic molecule acetylene. This establishes LIED as a methodology for dynamically visualizing larger and heteronuclear molecular structures. In this paper we use LIED to simultaneously measure C-C and C-H bond lengths of acetylene

($C_2H_2$) molecules. We chose acetylene as the test molecule since it is heteronuclear, readily alignable, linear and symmetric so that orientation is not required, and its bond lengths are accurately known. More importantly, however, is the fact that acetylene is a prototypical organic molecule in which the dynamics associated with isomerization, proton migration, internal vibrational redistribution of energy and conical intersections can be studied in the future using LIED. The molecules are either aligned along the polarisation direction of the laser field or anti-aligned, meaning perpendicular to the polarisation. The measured bond lengths lie within <5% of the expected acetylene cation equilibrium distances of 1.25 Å and 1.08 Å[28], respectively, for both molecular alignments.

The procedure for extracting $C_2H_2$ structural information using LIED is outlined in Fig. 2. Figure 2a shows the momentum distribution of all electrons detected in coincidence with all positive fragments after the ionization of $C_2H_2$ with our mid-IR source. The molecular differential cross section (DCS) is extracted by sweeping the scattering angle ($\theta_r$) around the circumference of a circle with radius equal to the momentum of the rescattered electron ($k_r$). The influence of the ionizing laser field must be considered, consequently, the origin of the circle is given by the vector potential ($A_r$) at the time of. Each circle represents rescattering by different electron energies. The extracted experimental molecular DCS ($\sigma_M$) is combined with the theoretical atomic DCS ($\sigma_A$) for the same electron energy and emission angle to calculate the molecular contrast factor (MCF) $M_F = (\sigma_M - \sigma_A)/\sigma_A$. The MCFs are typically presented as a function of the momentum transfer $q = 2k_r \sin(\theta_r / 2)$ experienced by the rescattered electrons. A $\chi^2$ based fitting routine is used to compare the experimentally obtained MCF to theoretical predictions.

The coincidence detection capability of the ReMi is crucial for accurate retrieval of polyatomic molecular structure from the experimental MCF. To develop the time resolving capabilities of LIED, it is important that we ensure the scattering pattern originates from the fragmentation channel of interest only. To highlight this point we present the time-of-flight (TOF) spectrum of all the detected positively charged fragments in Fig. 2b. The main peak near 4.2 µs is the acetylene cation ($C_2H_2^+$) investigated in this manuscript, and it constitutes about 10% of the total number of detected fragments. The inset shows a close-up of this peak and the black shaded area represents the region that the electrons associated with $C_2H_2^+$ are extracted from. Many other fragments can be observed and identified in the TOF and each of these peaks has associated electrons. Figure 2c shows the measured electron kinetic energy spectrum for all fragments (blue) and for the $C_2H_2^+$ fragment only (black). An order of magnitude difference in the number of detected electrons is visible over the entire spectral range. It is these omnipresent extra electrons that serve as an unwanted background signal and are detrimental to structure retrieval without coincidence detection. Figure 2d summarises this decisive point by comparing the MCFs retrieved when analysing electrons corresponding to all fragments (blue) and from $C_2H_2^+$ only (black). The $C_2H_2^+$ data results in an MCF that compares well with the equilibrium acetylene structure, which indicates that, in the case of acetylene, launching the recollision electron does not cause detrimental differences between the neutral and ionic species within the short recollision time. On the other hand, using electrons from all fragmentation channels results in a dramatically different MCF that cannot be accurately fitted and fails in retrieving the $C_2H_2^+$ bond lengths. We validate with this analysis that electron-ion coincidence detection is a pre-requisite for the application of LIED to larger molecules. The high sensitivity of LIED to the exact molecular structure is also illustrated in Fig. 2d by the dramatic change induced in the MCF by a 10% contraction (green) or expansion (red) of the molecule.

To visualize complex molecules, LIED needs to be able to retrieve multiple bond lengths between different atomic species. We demonstrate that our implementation of LIED fulfills this promise and is even able to image the (typically) elusive hydrogen atom by exploiting the fact that we measure the full doubly differential cross section. At the tens of keV electron energies used in UED a hydrogen atom has a scattering cross section ($\sigma_H$) that is typically much less than that of a C atom ($\sigma_C$). Figure 3a presents the $\sigma_H/\sigma_C$ ratio[29] for a 25 keV electron as a function of the scattering angle (green curve). The ratio is maximal for scattering angles between 0-5° that are typical in UED (shaded region), however, even in this range $\sigma_H / \sigma_C < 0.05$. The electron energies used in LIED, however, result in much higher values of the $\sigma_H / \sigma_C$ ratio, as is also presented in Fig. 3a for 50 eV (red curve) and 100 eV (blue curve) electrons, due to a minimum in the C atom differential cross section. Both of the presented energies have wide angular regions where $\sigma_H / \sigma_C > 0.10$ and a peak of $\sigma_H / \sigma_C > 0.50$ is observed near 80° for 50 eV electrons. The shaded regions represent the much wider scattering angles for which the LIED technique is valid.

To take advantage of the favourable cross-section ratio available to LIED, we confirm that we can simultaneously measure both the C-H and C-C bonds. The MCFs extracted for both molecular alignments after scattering of 60 eV electrons are presented in Fig. 3b. For aligned molecules (blue squares) the best theoretical fit (dashed blue curve) from the $\chi^2$ fitting routine results in bond lengths of $D_{CC}^{A,60} = 1.28 \pm 0.13$ Å and $D_{CH}^{A,60} = 1.04 \pm 0.10$ Å while for anti-aligned molecules (red circles) the same procedure results in estimates of $D_{CC}^{AA,60} = 1.33 \pm 0.13$ Å and $D_{CH}^{AA,60} = 1.15 \pm 0.12$ Å. Here the notation $D_{bond}^{alignment,energy}$ is used for the results of the individual fits to refer to energy-specific bond lengths. The estimated bond lengths agree well with the known values[28] and the accuracy of each fit is about 10 pm, which is an order of magnitude lower than the de Broglie wavelength of the scattering electrons ($\lambda_E = 1.3$ Å). The positions of the MCF extrema and zero crossings, as well as the peak to peak modulation, are very sensitive to changes in the bond lengths and the molecular alignment. It is the sensitivity of these parameters that is utilised to monitor sub-Ångström changes in molecular structure. The two MCFs presented in Fig. 3b show some differences such as the position of the minimum near $q = 3.5$ Å$^{-1}$, which is closer to zero for anti-aligned acetylene, and the modulation amplitude, which is smaller in the aligned case. Depending on the target and the degree to which it is aligned, molecular alignment or anti-alignment can lead to larger differences in the peak-to-peak amplitude of the MCFs, which is beneficial for structural imaging. These results confirm that LIED can simultaneously extract multiple bond lengths from complex polyatomic molecules with high accuracy.

Next, we illustrate the possible attosecond temporal resolution[30] of the technique in Fig. 4. We measure the doubly differential cross section, which permits retrieving the C-C (C-H) bond length as a function of the rescattering electron energy. Based on operating mid-IR LIED in the quasi-static limit we can invoke the classical rescattering model to associate a specific time to the measured electron rescattering energy. The top axis in Fig. 4 shows the corresponding return time for each electron energy and indicates that a temporal resolution below 100 as could be achieved by analyzing at different rescattering energies. We further elaborate that the measured energy range can also be used to establish an unprecedented level of confidence and redundancy for the retrieved bond length. The extracted $D_{CC}^{AA}$, $D_{CH}^{AA}$, $D_{CC}^{A}$ and $D_{CH}^{A}$ values are consistent with the estimated ionic equilibrium values (dashed black lines)[28] over the investigated energy range. As no significant structural rearrangements are expected after acetylene is ionized from a neutral to a cation[28] fitting a horizontal line to the energy dependent bond length estimates will yield an overall estimate of the C-C and C-H bond lengths. This fitting results in estimates of $R_{CC}^{AA} = 1.24 \pm 0.04$ Å and $R_{CH}^{AA} = 1.10 \pm 0.03$ Å for anti-aligned molecules while the same analysis with aligned molecules results in bond lengths of $R_{CC}^{A} = 1.26 \pm 0.04$ Å and $R_{CH}^{A} = 1.05 \pm 0.03$ Å. This method amounts to performing two dimensional fitting over both electron energy and scattering angle, which is not possible with other techniques, and highlights the accuracy of the LIED method.

In summary, we demonstrate a robust method for the retrieval of multiple bond lengths from an aligned polyatomic molecule with mid-IR LIED, which we validate by accurately determining the structure of acetylene. The use of a reaction microscope in combination with a home-built 160 kHz mid-IR OPCPA exploits coincidence detection together with the measurement of the doubly differential elastic scattering cross section in the quasi-static regime. This unique capability enables imaging the hydrogen atom by selection of a suitable scattering energy range for which the relative cross sections contribute comparably. An excellent bond length confidence level is achieved due to the large range of rescattering energies for which structural information is measured. Our method demonstrates a clear path to exploit the intrinsic atto- to femto-second temporal resolution of LIED for the imaging of complex molecules. Finally, we provide a solution to selective imaging of the multiple fragmentation pathways that are inherently created when launching the recollision electron in polyatomic molecules. This capability is an enabling step towards time resolved imaging and permits accurate retrieval of the geometrical structure from the fragment of interest only. Our data already contains structural information on the ultrafast isomerisation of acetylene and on multiply charged ions, which we aim to investigate in future work. The technique provides an accessible and robust route towards probing ultrafast processes in complex gas-phase molecules by combining attosecond and collision physics towards realising the molecular movie.

**Acknowledgements** We acknowledge support from MINISTERIO DE ECONOMA Y COMPETITIVIDAD through Plan Nacional (FIS2011-30465-C02-01), the Catalan Agencia de Gestió D'Ajuts Universitaris i de Recerca (AGAUR) with SGR 2014-2016, Fundacio Cellex Barcelona, and funding from LASERLAB-EUROPE, grant agreement 228334.

**Author contributions** J. B. conceived the experimental investigation. M. G. P., B. W., M. B. and M. H. acquired the experimental data. A. T. L. and C. D. L. provided theoretical support. A. S., C. D. S., J. U. and R. M. provided experimental support. M. G. P., B. W., A. T. L., C. D. L. and J. B. wrote the manuscript.

**Competing Interests** The authors declare that they have no competing financial interests.

**Correspondence** Correspondence and requests for materials should be addressed to Michael G. Pullen at michael.pullen@icfo.eu.

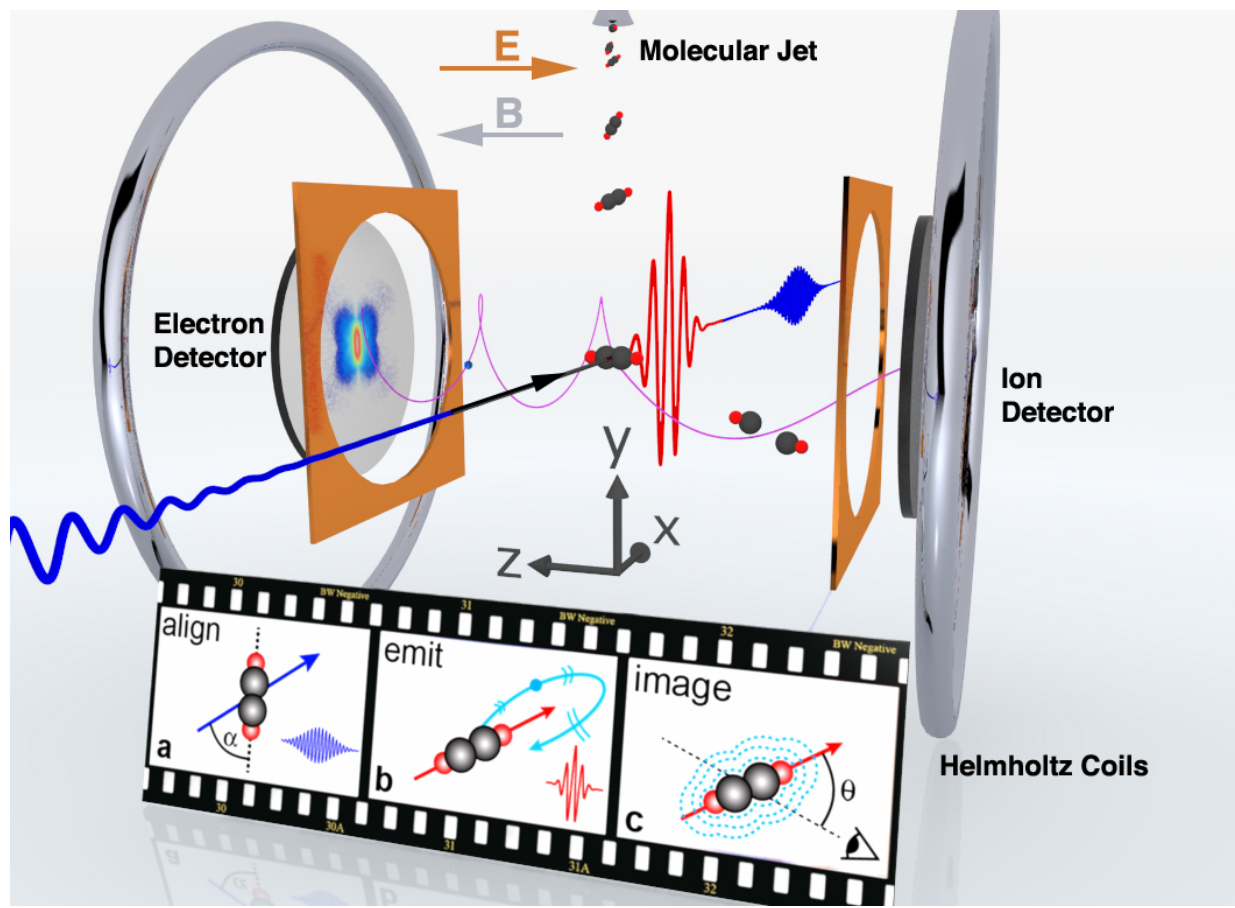

**Fig. 1. Laser-induced electron diffraction from aligned $C_2H_2$ molecules using a mid-IR OPCPA source and a reaction microscope.** The cartoon film shows the procedure: (a) The $C_2H_2$ molecules are pre-aligned by focusing the 1.7 um pump pulse (blue) into a molecular jet. (b) The 3.1 um pulse (red) is used to generate high energy electrons that subsequently rescatter off the parent ion. (c) The rescattered electrons carry structural information of the parent ion that is contained in the detected angular momentum distributions. The anti-collinear electric ($\vec{E}$) and magnetic ($\vec{B}$) fields guide the charged fragments towards opposing position sensitive detectors.

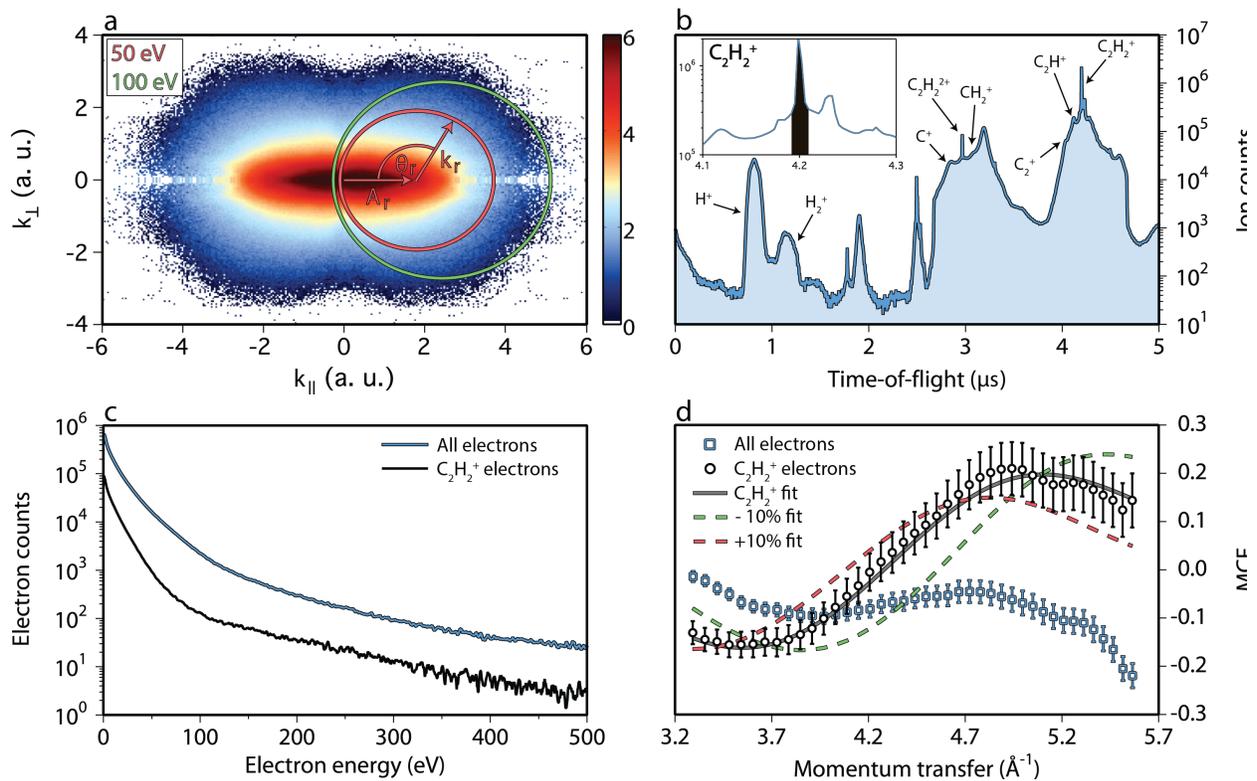

**Fig. 2. Method to extract structural information from the momentum distributions.** (a) Logarithmically scaled momentum distribution of electrons corresponding to all ionic fragments. The circles represent the scattering of electrons with the same energy at different angles. (b) The detected ion time-of-flight (TOF) showing the numerous fragments created during the strong-field interaction. The inset shows the peak corresponding to the $C_2H_2^+$ ion near 4.2 μs and the shaded region represents the window that the $C_2H_2^+$ electrons are taken from. (c) The electron kinetic energy distribution for the $C_2H_2^+$ ion (black) and for all possible fragmentation processes (blue). (d) The extracted MCF for the acetylene cation (black circles) as well as for electrons from all fragments (blue squares). The solid black curve shows the best fit, which matches very well with the cation channel. The MCFs for ±10% changes in the $C_2H_2$ molecular lengths (dashed curves) highlight the sensitivity of the LIED technique.

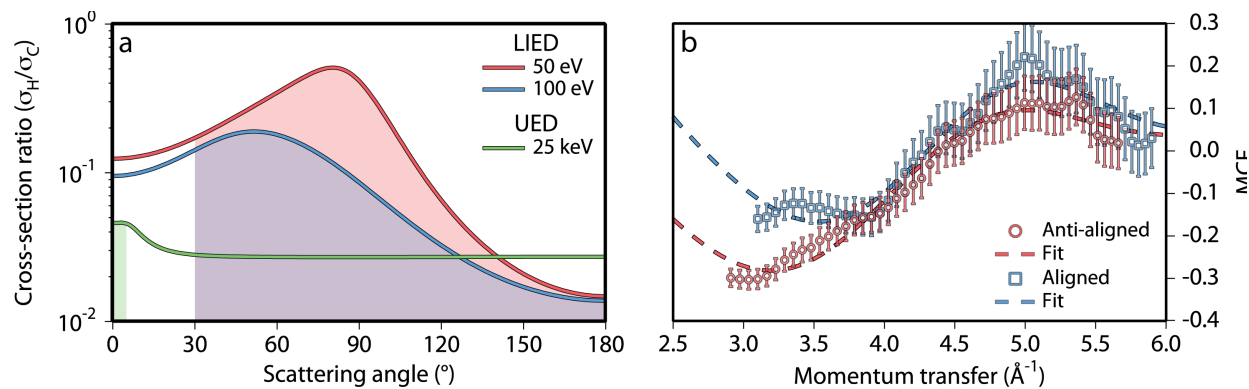

**Fig. 3. Simultaneous extraction of multiple bond lengths from polyatomic molecules.** (a) The ratio of the H and C scattering cross-sections as a function of electron scattering angle for typical energies used in LIED (50 & 100 eV) and CED/UED (25 keV). The ratios are much higher for the energies relevant to LIED and are also applicable over a much wider angular range (shaded regions). (b) Blue squares (red circles) show the experimental molecular contrast factor from the scattering of 60 eV electrons off aligned (anti-aligned) molecules. The best theoretical fits (dashed lines) allow the accurate extraction of the C-H and C-C bond lengths from both alignments.

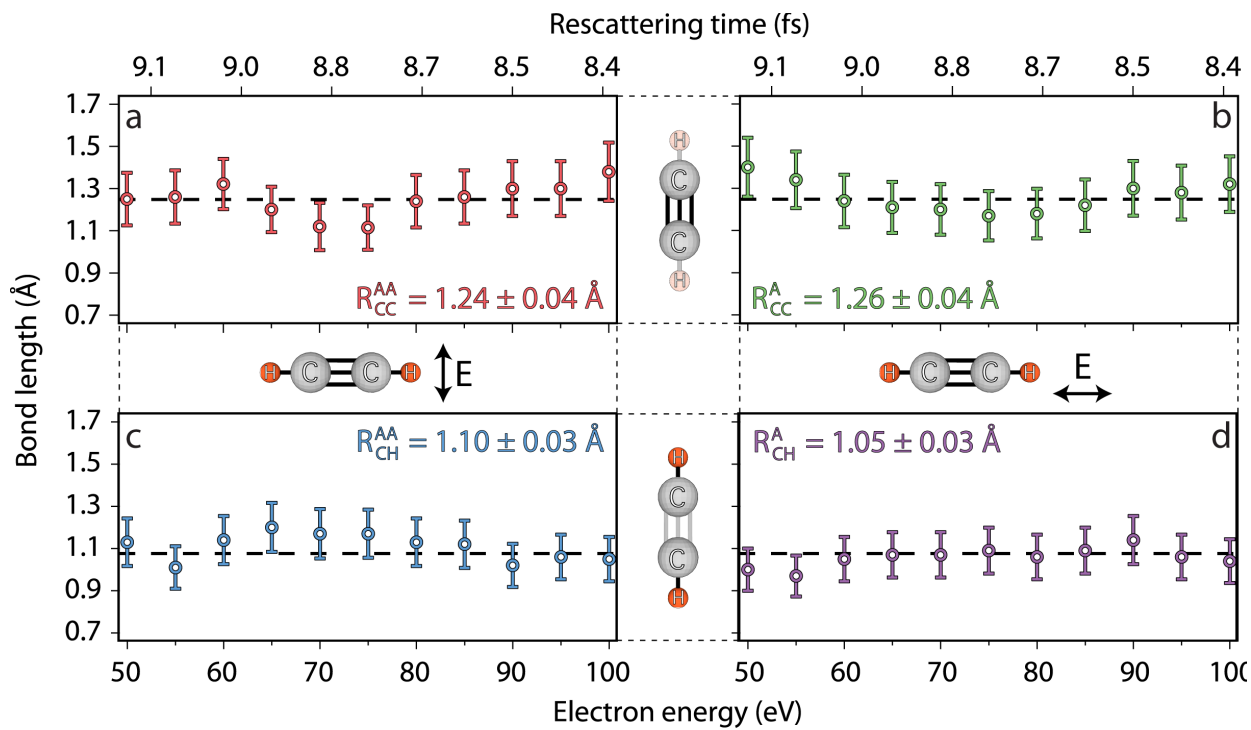

**Fig. 4. Accurate $C_2H_2$ bond length extraction.** The C-C (C-H) bond length estimates are presented as a function of the scattering electron energy and rescattering time in the top (bottom) quadrant. The expected equilibrium values of the acetylene cation are also shown (dashed black lines). The values of the best horizontal fits for each bond are displayed in the respective panels.